# BUTO FACULA, GANYMEDE: PALIMPSEST EXEMPLAR

Jeffrey M. Moore[1], Oliver L. White[1,2], Donald G. Korycansky[3], Paul M. Schenk[4], Andrew J. Dombard[5], and Martina L. Caussi[5].

[1]NASA Ames Research Center, Moffett Field, CA, 94035, [2]SETI Institute, Mountain View, CA, 94043, [3]University of California, Santa Cruz, CA, 95064, [4]Lunar and Planetary Institute, Houston, TX, 77058, [5]University of Illinois at Chicago, Chicago, IL, 60607.

**Abstract**

Nowhere in the solar system are impact morphologies observed in greater variety than on the icy Galilean satellites. This is very likely a consequence of the structural and thermal state of the crust at the time of impact, and perhaps impact velocity. The palimpsest class of impact feature displays smooth enclosed central plains surrounded by an expanse of undulating plains, within which are distributed concentric arcuate ridges, and no recognizable rim. Buto Facula is the best resolved of any palimpsest on Ganymede or Callisto, with the majority of it having been imaged by *Galileo* at 190 m/pixel and optimum lighting, permitting insight into the circumstances that form this type of impact feature. The western half of a pre-existing, 19 km diameter impact crater situated at the eastern edge of Buto Facula has been buried by material of the undulating plains. This suggests that at the time of their emplacement the undulating plains behaved as a low-viscosity flow advancing across the landscape that surrounded the impact zone, encroaching on landforms that it encountered. We evaluated several hypotheses for the formation of undulating plains using impact and ejecta modeling. A primary reason for not attributing the source of Buto's undulating plains to "dry" impact ejecta is the existence of impact features that are as large or larger than Buto, such as Gilgamesh, that are not surrounded by palimpsest-like undulating plains deposits. We performed iSALE impact simulations incorporating a subsurface liquid layer (or low strength layer) at various depths. An impact into a surface with a pre-existing, 5 km-thick fluid layer located at a depth of 5 km will result in excavation of fluid material from that layer, producing a nearly flat final surface profile that is consistent with the flat profile of Buto and the distribution of its undulating plains material. Moving the liquid layer to greater depths beyond a range of ~20-40 km results in final impact feature profiles that resemble classic impact craters. We offer several tests of this shallow subsurface liquid layer hypothesis for palimpsest formation that could potentially be performed by *JUICE* and *Europa Clipper*

## 1. Introduction

The Galilean satellites with icy surfaces (Europa, Ganymede, Callisto) are host, among other things, to a variety of large impact features that are rarely encountered on other planetary and satellite surfaces in the Solar System. These features include impact basins with central pits, domed floors, and so-called "penepalimpsests" and "palimpsests" (Passey and Shoemaker; 1982). The present authors have conducted a multi-disciplinary approach to study how these features formed (White et al., 2024; Caussi et al., 2024; Korycansky et al., 2022a, 2022b). It is likely that the particular combination of geophysical factors and impactor characteristics that is shared by these satellites is responsible for these features. One component of this study has been to derive the topography of a number of these features, and then use these Digital Terrain Models (DTM) (see Schenk et al., 2018; Schenk et al, 2004a) as a fundamental tool in producing facies maps of a range of these features for which adequate data exist (White et al., 2024).

White et al. (2024), building on Schenk et al. (2004a), recognized that impact features on Ganymede and Callisto fall into two broad but distinct classes: (a) crater forms and (b) palimpsests (including penepalimpsests, which henceforth will be referred to collectively as palimpsests in this study). Crater forms all share common characteristics, the most important of which is a recognizable crater rim. Larger crater forms exhibit common facies that are not encountered in smaller craters, which are either bowl-shaped simple craters or flat-floored complex craters with central peaks, morphologies that are universally encountered throughout the Solar System. The morphologies of these larger impact features appear to be size-dependent. The smallest of these morphologically distinct crater classes are pit craters, which are less than 100 km in diameter and feature a central pit on their floors surrounded by a raised annulus, or in some cases multiple small pits distributed within the raised annulus. Craters around 100 km in diameter display domes on the floors of their pits and so have been termed dome craters. Craters larger than ~110 km in diameter display very subdued and broad crater rims that do not form a sharp ridge as in smaller crater classes, although they still display prominent annuli, pits, and domes, and are described as anomalous dome craters (Schenk et al., 2004a). Anomalous dome craters can reach more than 250 km in diameter, and their size range overlaps with that of palimpsests, which White et al. (2024) interpreted to indicate that the impact size is less important in determining impact feature morphology at these large diameters.

Palimpsests (and features sometimes classified as penepalimpsests, e.g., Passey and Shoemaker, 1982; White et al., 2024) share a set of common facies that are very different to those shared by craters, and show very little topographic relief by comparison: typically a few hundred meters while craters can show more than a few kilometers (White et al., 2024). White et al. (2024) described palimpsests and penepalimpsests as being end-members of the same morphological class, with penepalimpsests displaying well-developed, arcuate ranges of concentric ridges or hills, while such ridges occur only sporadically or not at all for palimpsests. In this paper we have lumped both of these types together, referring to all of them as palimpsests. Figures A and Ba illustrate five palimpsests on Ganymede as geologically mapped by White et al. (2024). Palimpsests do not exhibit recognizable topographic (crater) rims. They all have a low-lying, smooth-to-finely textured expanse of Central Plains that are enclosed by a shallow, inward facing scarp. An Undulating Plains facies extends beyond the Central Plains, which makes up most of the surface of palimpsests.

The roughly circular outer contact of the Undulating Plains is abrupt where it is superposed on pre-existing terrain beyond the palimpsest, usually of a lower albedo. In some cases, secondaries from the palimpsest-forming impact can be recognized on the terrain beyond the outer edge of the Undulating Plains. Often found amongst the Undulating Plains are nested, arcuate Concentric Ridges, the only palimpsest facies with any appreciable positive topographic relief, and the prevalence of which varies between palimpsests. Palimpsests that are older (as determined by crater counts, Schenk et al., 2004a; White et al., 2024) and larger, such as Memphis Facula, tend to exhibit fewer Concentric Ridges than those that are younger and smaller. Here we present analysis of Buto Facula, Ganymede, as an exemplar of a palimpsest, including penepalimpsests (e.g., Passey and Shoemaker, 1982). Buto Facula is the best resolved of any palimpsest on Ganymede and Callisto, with the majority of it having been imaged by *Galileo* at 190 m/pixel, permitting insight into the circumstances that form this type of impact feature.

2. **Observations**

Buto Facula (located at 13.2°N, 203.5°W) displays an approximately circular expanse of rough, Undulating Plains ~240 km in diameter, within which rise nested concentric arcuate ridges (Fig. B). The Undulating Plains facies embays the concentric arcuate ridges so we infer that these ridges are blocks of the upper crust that were promptly inwardly rotated by slumping during and immediately following the impact event. At the center of Buto are the smooth Central Plains that are enclosed by a low, inward-facing scarp. This Central Plains material superposes the surrounding Undulating Plains, and is littered with small blocks reminiscent of "small chaos" on Europa (Greenberg et al., 1999). Schenk et al. (2004a) and White et al. (2024) have highlighted a well-defined, pre-existing 19.1 km diameter impact crater at the eastern edge of Buto Facula, of which roughly the western half has been mantled (or buried) by material of the Undulating Plains of Buto

Facula (Fig. C), suggesting that the Undulating Plains at the time of their emplacement may behave as a low-viscosity flow advancing across the landscape that surrounds the impact zone, encroaching on landforms that it encounters. We have been able to derive a good estimate of the thickness of the Undulating Plains material where it covers this crater by taking 12 radial profiles across the crater. In our profile plot in Fig. C the blue profiles are the six radial profiles taken across the crater's unmantled eastern half, and the orange profiles are the six radial profiles taken across the mantled western half. The thick blue profile is the averaged unmantled profile, and the thick red profile is the averaged mantled profile. The rim crest is much lower for the mantled portion than the well-preserved portion (by about 220 m), but the most informative difference in terms of how much Undulating Plains material is covering the crater is the portion of the mantled part of the crater interior between 10.8 and 17.5 km radial distance (indicated by the green outline in the map in Fig. C), which is much shallower compared to the unmantled side. We integrated between the averaged mantled and exposed profiles between 10.8 km and 17.5 km to obtain a total volume of deposited material within the green outline of 27.7 km$^3$, spread over an area of 288 km$^2$, which equates to a mean thickness of 96 m. Importantly, the profiles indicate that the Undulating Plains material is not spread equally across the underlying crater, as would occur for uniform accretion, but rather it concentrates in the annular depression that forms the outer portion of the crater floor. There may be little to no material covering some segments of the subdued western crater rim crest. As such, 96 m would represent a maximum thickness at this distance from the palimpsest center, although it would be expected to increase towards the center. The mapped area of Buto (excepting the Central Plains) is 40,610 km$^2$, equating to a volume of almost 3900 km$^3$ for the Undulating Plains assuming a universal thickness of 96 m.

    We infer that the surface morphology of the Undulating Plains exhibits characteristics that indicate the material composing this unit behaved similar to a liquid, slurry, or some other

mechanically weak material at the time of emplacement. Three main hypotheses have emerged to explain the morphologies of palimpsests: firstly, that the margin of the palimpsest coincides with the rim of the original crater (e.g. Hartmann, 1984); secondly, that the rim of the original crater lies well inside the margin of the palimpsest, which represents the limit of continuous crater ejecta (e.g. Passey and Shoemaker, 1982; Schenk et al., 2004a); and thirdly, that palimpsests result from protracted volcanic extrusions triggered by large impact events early in Ganymede's history, with impact excavation occurring to a sufficient depth to penetrate the thin, primordial lithosphere, allowing mobile, buoyant material from a subsurface liquid layer to rise to the surface (Greeley et al., 1982; Thomas and Squyres, 1990). In this paper we apply impact and ejecta modeling to the problem in order to test these hypotheses, using our measurements of the palimpsest profile and the volume of the Undulating Plains material to evaluate our results.

## 3. Impact Modeling:

Supplementing these observations and measurements, impact simulations and crater scaling relations can perhaps shed further light on the formation of Buto Facula, or at least provide useful constraints and tests of our hypothesis.

## 3.1 Impact ejecta

The first question we look at is where continuous impact ejecta originates from in the target and its volume. In our numerical simulations, we have results from running the iSALE code (Amsden *et al.*, 1980; Collins *et al.*, 2004; Wünnemann *et al.*, 2006). The specific version we used is "iSALE- Dellen" (Collins *et al.*, 2016). Both impactor and target are pure ice, with the standard ANEOS equation of state for $H_2O$ that is included with the iSALE distribution. Simulation resolution in the central portion of the grid was 18 cells per impactor radius, i.e. $\Delta r = \Delta z = 200$ m. The target temperature is 120 K. For an impactor diameter $d_i = 7.2$ km, an impact velocity $U_i = 20$ km s$^{-1}$, and a vertical impact, we have at the end of the calculation a crater profile as shown in Fig D. We determine the surface $z_s(r)$ as a function of radius, where the surface is the grid cell in which density transitions from ~0 to ~$\rho_{ice}$. The surface profile crosses the $z = 0$ plane at radius $R_0 = 38.6$ km, and the peak of the rim falls at radius $R_p = 44.6$ km. The volume of the impact crater $z < 0$ is $2.3 \times 10^4$ km$^3$.

We attempt to evaluate ejecta volumes by means of Lagrangian tracers (a capability of iSALE) that are embedded in the target on the grid at the start of a calculation. The tracers have no inertia of their own, but get their instantaneous velocities by interpolation from the grid, followed by positional updates using those interpolated velocities. Tracers are ascribed pressures, densities, and temperatures interpolated from the grid cells in which they are located. Each tracer is labeled with an index $i$, and located at position $r_i, z_i$. Additionally, each tracer has an interval $\Delta r$ and $\Delta z$ that are the same, as the tracers are equally spaced. We associate a specific volume with each tracer based on the interval $\Delta r$ and $\Delta z$ that specifies their separation in the target. The volume associated with a tracer $i$ located at $r_i, z_i$ is

$\Delta V_i = \pi(r_{i+1/2}^2 - r_{i-1/2}^2)\Delta z$, where $r_{i-1/2} = r_i - \Delta r/2$, and $r_{i+1/2} = r_i + \Delta r/2$.

Note that the code grid is Eulerian and is fixed, but tracers are Lagrangian and move through the fixed grid. We assume that $V_i$ is preserved no matter where tracer $i$ goes during the calculation. Therefore, the ejecta volume from the impact is the sum of volumes $V_i$ for tracers $i$ that fulfill specific criteria at the end of the calculation. The final positions of such tracers (as of 850 s after the impact) have 1) radii $r_i > R_p$ and 2) lie within 500 meters depth of the surface are counted as part of the ejecta. We may expect that there will be a fraction of material that is melted or vaporized that would otherwise produce ejecta near the crater. The fact that the impactor and target are made of ice presents a complication in the form of melting and/or vaporization of material from the target. We therefore exclude tracers for which densities fall below $\rho = 10^2$ kg m$^{-3}$ from our ejecta count. A plot of the paths of tracers that we use to assess ejecta volume from the impact of an object of diameter $d_i = 7.2$ km and $U_i = 20$ km s$^{-1}$ is shown in Fig. D. Tracers shown in Fig. D fall within ~90 km of the impact center, somewhat closer than the average radius of Buto Facula of ~120 km.

A feature of iSALE that is relevant here is the so-called density cutoff that is used in the code (Collins *et al.*, 2016). Material in grid cells with densities less than the cutoff value is eliminated from the grid in order to prevent situations in which the specific energy of low-density material spuriously increases to unrealistically high values, causing a "timestep crash" that halts the calculation. The cutoff density parameter (denoted ROCUTOFF in the input file) is user-definable, with a default value of 5.0 kg m$^{-3}$.

A side effect of the density cutoff is that tracers that also might be present in density-cutoff cells are also removed from the calculation. In order to attempt to track ejecta better we carried out an impact simulation with ROCUTOFF set to $1 \times 10^{-6}$ kg m$^{-3}$; a timestep crash occurred at $t = 856$ s after the impact, but that was sufficiently far enough along to track the re-impact of ejecta with the surface outside the crater.

Using the outcome of a hydrocode simulation, we infer a volume of ~8000 km$^3$ for the ejecta, which is larger than the estimated volume of the Undulating Plains unit at Buto Facula by approximately a factor of 2, but as we were only able to determine its thickness at the distal edge, we suspect that this results in an underestimate of the total actual volume of the unit. Regardless of the quantity of ejecta produced, the final surface profile of the impact feature generated in our ejecta simulation in Figure D is distinctly bowl-shaped with several km of relief, and therefore is in no way similar to the flat and rimless Buto Facula. The whole process of excavation must be different to that of other large impact features on Ganymede that show crater-like morphologies. We therefore must seek another explanation for Buto Facula's Undulating Plains besides ejecta deposited from an impact into a solid ice target.

### 3.2 Surface profiles from impact simulations

We turn to calculations designed to produce surface profiles that are more aligned with that of present-day Buto Facula. The surface profiles shown above in Figs. D and E were produced from an impact into a "default" homogeneous ice target with constant temperature ($T$ = 120 K) as a function of depth and a default strength.

The model for material strength and damage is given by the ROCK and Ivanov damage models defined in iSALE (Collins et al 2016). The parameters for the ROCK strength model were given values listed in Table 1. The values that we chose for the parameters are typical for general models of impacts into ice targets found in the outer solar system.

We can make a distinction between the issues of surface profiles and the Undulating Plains mentioned previously; in principle a profile that matches that of Buto Facula might not entail the concomitant formation of a surface layer like the Undulating Plains. Nonetheless the calculations described in this section also seem to produce a thin layer of melt on the surface, which might match the observed geological feature.

We investigated the effects of impacts on targets with a subsurface layer, either of melt (i.e. a layer of temperature $T = T_{melt} = 273$ K) or of substantially reduced material strength. The iSALE code is capable of including such effects in its target specifications. For these calculations we return to the default value of the density cutoff ROCUTOFF = 5 kg m$^{-3}$.

Turning first to the effects of a pre-existing liquid layer, for definiteness we posit a liquid layer of 5 km thickness at various depths. We look at cases where the bottom of the liquid layer is at depths of 10, 20, 40, or 60 km. A plot of the 10-km case is shown in Fig. F. Surface profiles from the various simulations are shown in Fig. G. As might be expected the shallowest liquid layer has the largest effect, producing a virtually flat surface. Comparison of the results for deeper layers shows the diminishing effect of those deeper layers on the surface profile. We compare these profiles to an averaged radial profile of Buto, generated from 10 raw radial topographic profiles taken across the DTM in Figure Bb. We note that since the radial spacing of concentric hills with changing azimuth around the palimpsest is irregular, the topographic signatures of the concentric hills and the Undulating Plains merge in the averaged profile – the reliefs of individual hill ranges are apparent in the raw profiles and can reach up to ~300 m above the Undulating Plains. Given that the overall topographic relief of the averaged profile does not exceed a few hundred meters, and displays no increase in surface elevation from the center of the palimpsest towards its edges, it is most similar to the simulated profile with a liquid layer at a depth of 10 km.

Two caveats apply for attributing the profile of Buto Facula (or other palimpsests) to the effects of pre-existing subsurface liquid layers. First, we have not taken into account the effects of freezing a pre-existing subsurface liquid layer. As is well known, upon freezing, water ice expands by approximately 10%. Likewise, we have not made any estimate of the effects of viscous relaxation on surface profile (e.g,. Caussi et al, 2024). Doing so would require modeling the long-term evolution of the surface, which we have not done here. A second caveat is that any near sub-surface liquid $H_2O$ layer would be expected to freeze, perhaps on a short timescale relative to that required to generate the population of palimpsests. While we will not go into detail here, a one-dimensional calculation that includes latent heat of the freezing of a 5-km thick layer of melt sandwiched between ice layers gives a timescale of ~ $10^6$ years for freezing. Possible explanations for a longer lived near sub-surface liquid layer are the layer is (1) a eutectic mix of $H_2O$ and $NH_3$, (2) a $H_2O$ mixture of significant $H_2SO_4$, or (3) a concentrated brine, in combination with higher inertial heat flow and an insulating surface cover.

We have also run calculations with weak layers where the ice temperature in the layer is the same as in the surroundings (120 K) but the material strength has been reduced by 10 or 100 times from the nominal values. Effects for the same layer placements as before for the 100× weaker layer are shown in Figs. H and I. In this case we see that the effects of a weak layer are qualitatively similar to a melt layer, but in general only the shallowest weak-layer case produces a flat surface profile. Again, it should be noted that we have not *as yet* modeled the further long-timescale evolution of the formation, although a weak layer in thermal equilibrium with its surroundings should be able to persist on geological timescales.

4. **Discussion and Conclusions**

The volume of Undulating Plains material that we have measured in this study is substantially less than the amount of target material that would have been excavated by the impact event itself according to our simulations of ejecta production by an impact into a solid ice target. We also point out an additional reason for not attributing the source of the Buto surface layer to "dry" impact ejecta. This is simply the existence of all the other large craters on Ganymede and Callisto that are not surrounded by palimpsest-like "Undulating Plains" deposits. This includes Gilgamesh, which is a relatively recent impact event that is comparable to or larger in size than palimpsests and does not have associated Undulating Plains (Collins et al., 2013). Considered alongside the finding that the very low topographic relief and absence of a rim exhibited by Buto's Undulating Plains is replicated in our impact simulation with a shallow pre-existing liquid layer, we conclude that the Undulating Plains material was instead mostly derived from a liquid or slurry layer in the near-surface target at the time of impact, in agreement with several other studies (e.g., Greeley et al., 1982; Thomas and Squyres, 1990; Schenk et al, 2004a). The impact that created Buto penetrated into this subsurface liquid layer through a thin overlying ice layer, liberating thousands of cubic km of this liquid that ascended to and spread across the surface to form the Undulating Plains, with only a minor contribution from impact melt.

As argued by White et al. (2024), the existence of the Concentric Ridges at Buto and other palimpsests indicate that liberation of liquid from this near-surface layer did not completely define the appearance of the final impact feature, with portions of the overlying surface ice layer being rotated, but otherwise preserved in solid state, to form these ridges. An interesting aspect of Buto is that, while the Concentric Ridges of other palimpsests appear essentially circular in planform, those of Buto form a squared-off outline that appears to align with a northeast-southwest-oriented structural fabric that characterizes this region. Existing regional jointing has been hypothesized to affect the morphologies of craters at very different scales on other planetary bodies (e.g., Shoemaker and Kieffer, 1979), and Buto demonstrates that even for a palimpsest that is formed primarily through the mobilization of shallow liquid in the target, its final morphology can still be subject to structural anisotropies in the overlying ice shell.

The recognition of the Undulating Plains as the dominant unit in Hathor, Teshub, Nidaba, Zakar, Epigeus, and Memphis Facula, the other palimpsest-like features mapped by White et al.

(2024) (which display highly variable crater counts, with Buto being the least cratered amongst those palimpsests that are well-resolved) leads us to conclude that these features only form where the impact is large enough to penetrate through to a subsurface liquid layer. Where the ice shell is thicker, larger impacts are necessary to penetrate through the shell to form a palimpsest. By extension, all other impact features on Ganymede and Callisto, such as those with central pits and central domes, formed in targets with no or, at most, minor amounts of pre-existing liquid. Supporting this conclusion, recent modeling (Caussi et al., 2024) indicates that the present topography of impact features with central pits and domes can be entirely explained by the behavior of ice. Our study supports explanations for the formation and final appearance of palimpsests (and those features sometimes referred to as penepalimpsests) that are entirely applicable to this feature class, and sharply distinct from explanations for other impact features on Ganymede and Callisto.

Instruments aboard ESA's *Jupiter Icy Moons Explorer* (*Juice*) mission, with its dedicated orbital global examination of Ganymede (e.g., Stephan *et al.*, 2021), and NASA's *Europa Clipper* mission, which is likely to make several (non-targeted) flybys of Ganymede, hold great promise to test our hypothesis. *Juice*'s laser altimeter (Enya *et al.*, 2022) should produce topographic maps of palimpsests/penepalimpsests at high quality vertical and horizontal resolution, which will allow characterization of the thickness and volume of Undulating Plains units of virtually all these features on Ganymede. This will permit a substantial improvement in the quantitative characterization of these deposits. Laser Altimetry and high-resolution imaging (Della Corte *et al.,* 2014; Turtle *et al.,* 2024) will likely reveal features and textures on Undulating Plains that can be used to evaluate the rheological state of this material at the time of emplacement. Ground penetration radar (RIME on *Juice*) (Bruzzone *et al.*, 2013) should permit a determination of the thickness of Undulating Plains as well as potentially reveal evidence of the formerly liquid sub-surface layers, which we hypothesize were the impact-excavated source of Undulating Plains material. There might be fortuitous data acquired by Clipper's REASON (Blankenship *et al*., 2024) ground penetration radar at Ganymede, which operates at two frequencies (9 and 60 MHz) and could significantly supplement RIME (9 MHz only) observations. Mapping IR Spectrometers (MAJIS on *Juice*, MISE on *Clipper*) (see Poulet *et al*., 2024; Blaney *et al*., 2024 respectively) can potentially recognize compositions unique to Undulating Plains and hence reveal why this material might have been able to remain liquid during the era of palimpsests/penepalimpsest formation.

A final note is the potentially great importance of being able to relate similar observations (ideally with the same instruments) of the multi-ringed impact features (Tyre and Callanish) on Europa to palimpsests/penepalimpsests. Tyre and Callanish exhibit many of the same facies as do palimpsests/penepalimpsest on Ganymede. In particular, these Europan impact features have units analogous to scarp-enclosed Central Plains surrounded by large outer annuli of Undulating Plains, in which are situated Concentric Ridges (e.g., Moore *et al.,* 2001). The two principal differences are that Tyre and Callanish have concentric graben and are roughly a factor of three smaller than Ganymedean palimpsests/penepalimpsest. In as much that there is great interest in determining whether the Tyre and Callanish impacts completely penetrated Europa's ice shell, or not (e.g., Singer, *et al*., 2024), this determination may rest largely on a comparative understanding of these features on their respective satellites.

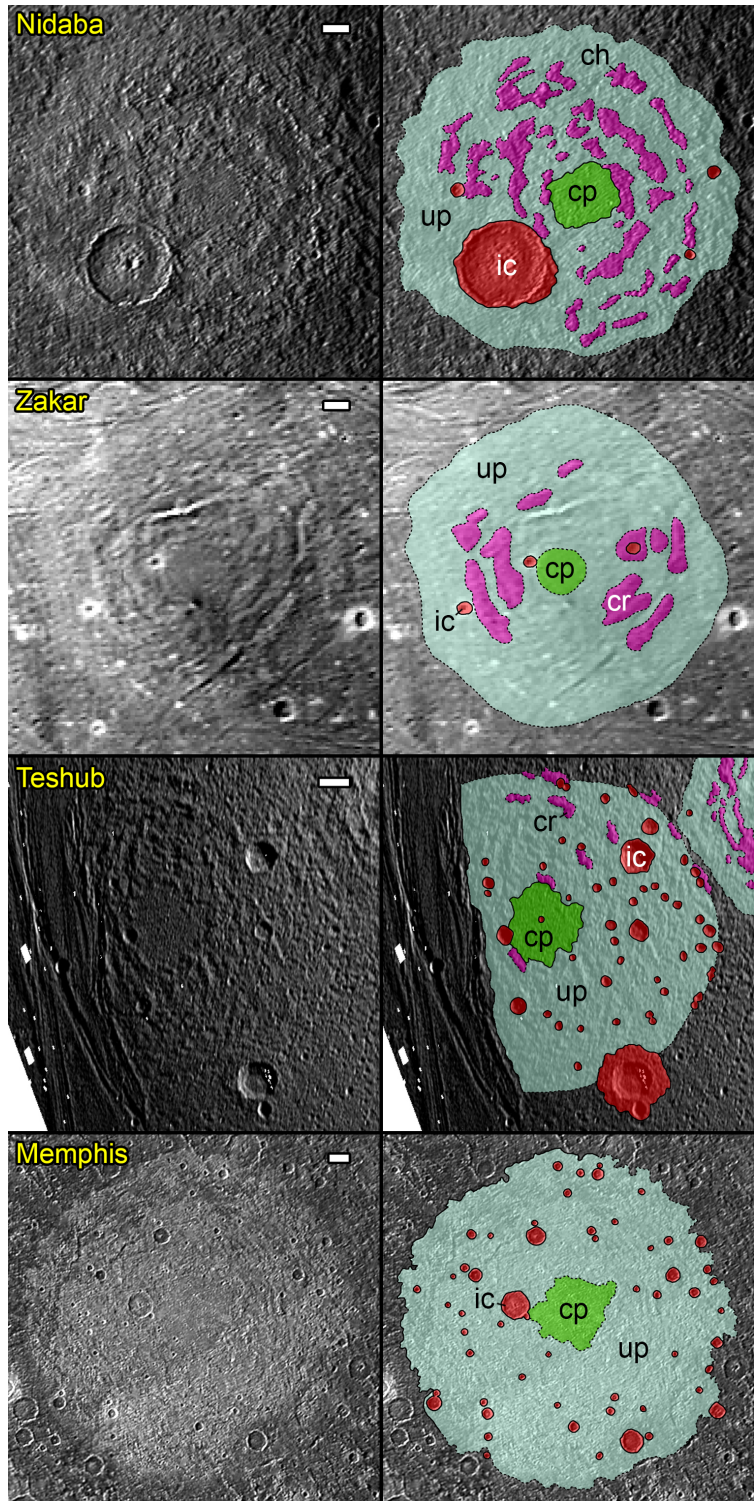

**Figure A.** Facies commonality of palimpsests (here including "penepalimpsests). Geologic maps modified from Fig. 3 in White et al. (2024) are shown at right, and the imaging used as the base maps for mapping shown at left. They all have: Central Plains (cp, green), which are low-lying and enclosed by a shallow, inward-facing scarp; Concentric Ridges (cr, pink), except the apparently oldest and largest palimpsests (e.g., Memphis Facula); Undulating Plains (up, cyan), which make up most of the surface; and superposing Impact Craters (ic, red). Palimpsests have no recognizable rim and the outer boundary of the Undulating Plains is indicated by an abrupt transition to exterior terrain of rougher relief and often lower albedo. White scale bar = 20 km in all maps.

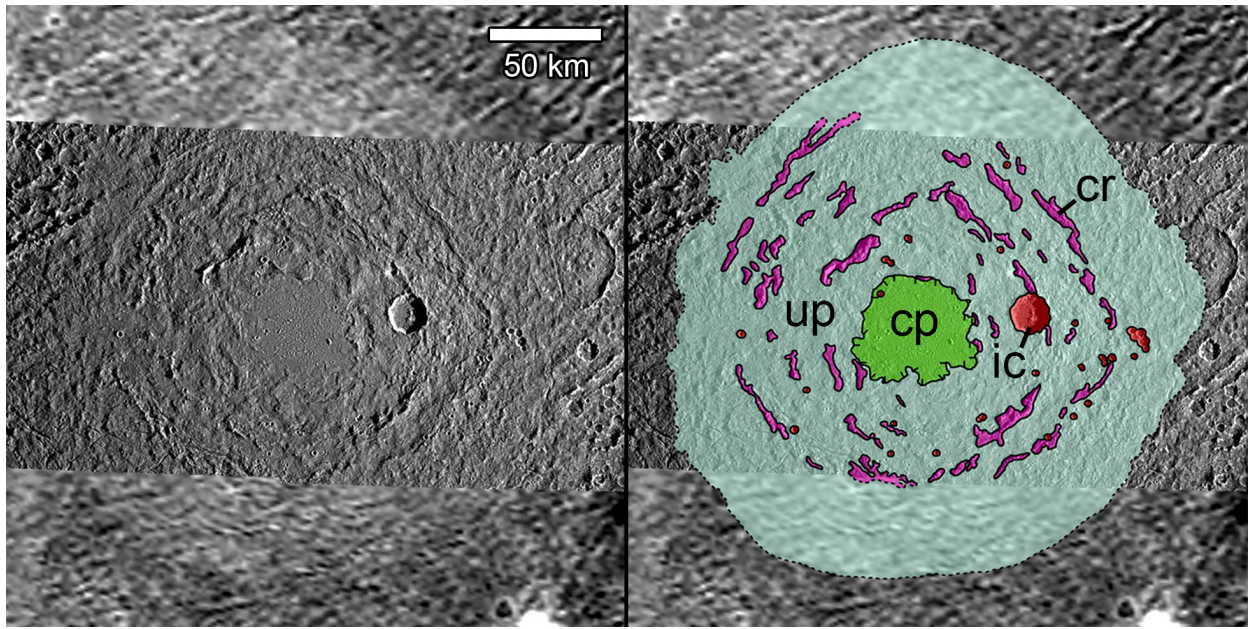

**Figure Ba.** Base map and geologic map of Buto Facula, as shown in Fig. 3 of White et al. (2024). Mapped facies and associated labels are the same as those in the legend of Fig. A. (Galileo observation G8GSBUTOFC01. North is up. Resolution 180 m/pixel. Incidence angle is ~80°. Background is Voyager 1 image FDS 20635.45. Resolution 1.5 km pixel.)

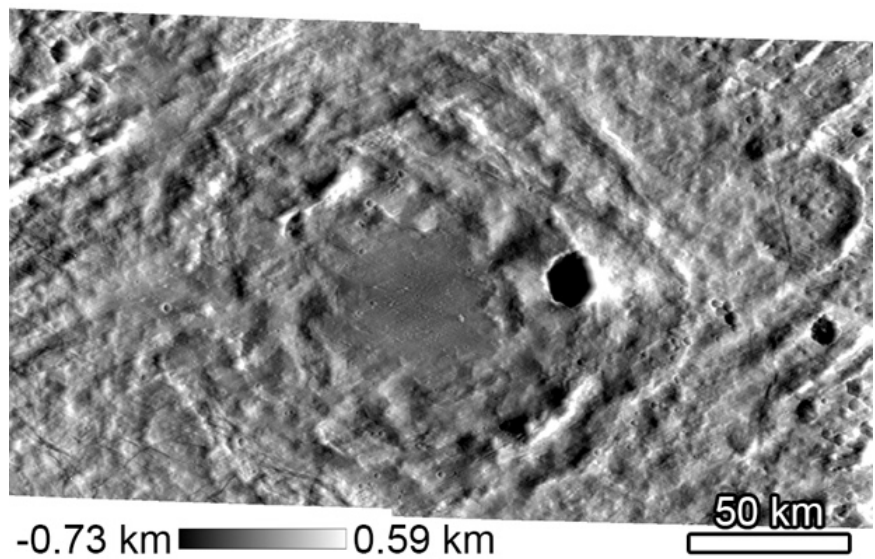

**Figure Bb.** DTM of Buto Facula generated from photoclinometry applied to Galileo SSI images of Buto taken under optimal lighting conditions for this technique (Schenk et al., 2004b).

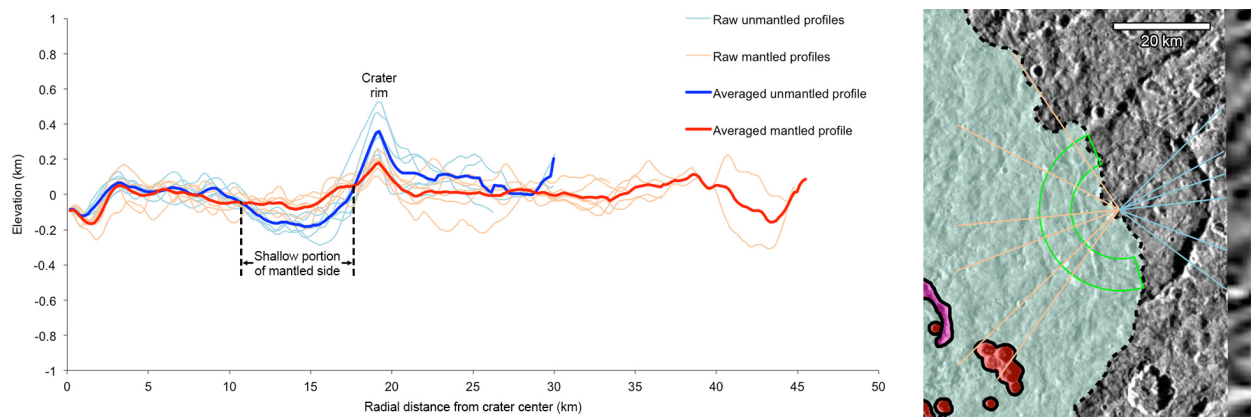

Figure C. Topographic profiles taken across a crater on the eastern edge of Buto that is partially mantled by material of the Undulating Plains unit, as shown in the detail of the geologic map in the right portion of the figure. Locations of the twelve profiles taken across the mantled and unmantled portions of the crater (six profiles for each) are shown on the map as orange and blue lines respectively. Green outline indicates the area on the floor of the crater for which we have calculated a volume of Undulating Plains material filling the crater using the averaged mantled and unmantled profiles.

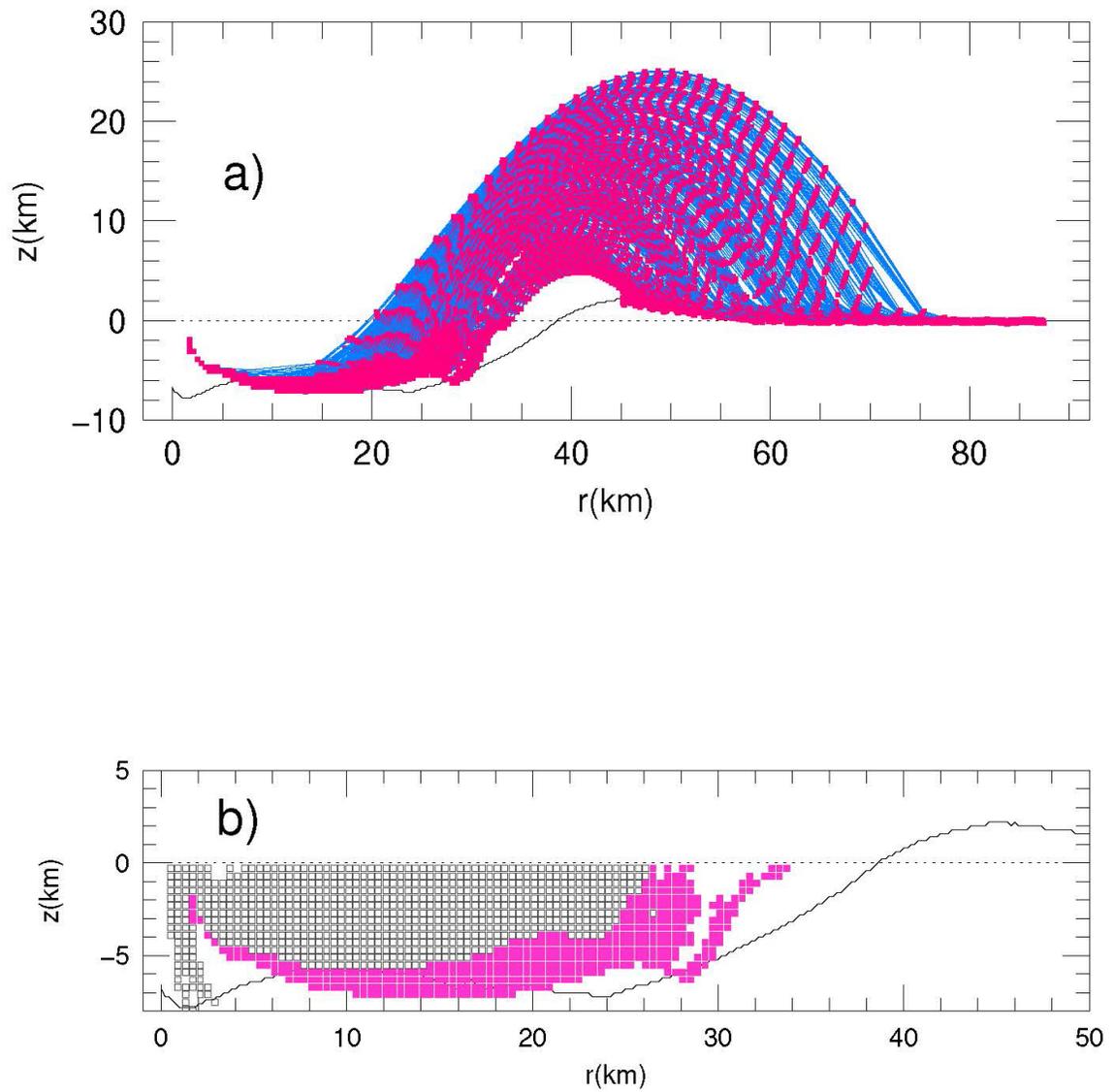

Figure D: a) Plot of paths of tracers that we use to assess ejecta volume from the impact of an object of diameter $d_i$ = 7.2 km, $U_i$ = 20 km s$^{-1}$. Tracer paths shown in blue, connecting tracer positions in purple. Final profile of crater shown in black. b) Plot of initial condition of tracers; purple ones are those that were tracked in panel a).

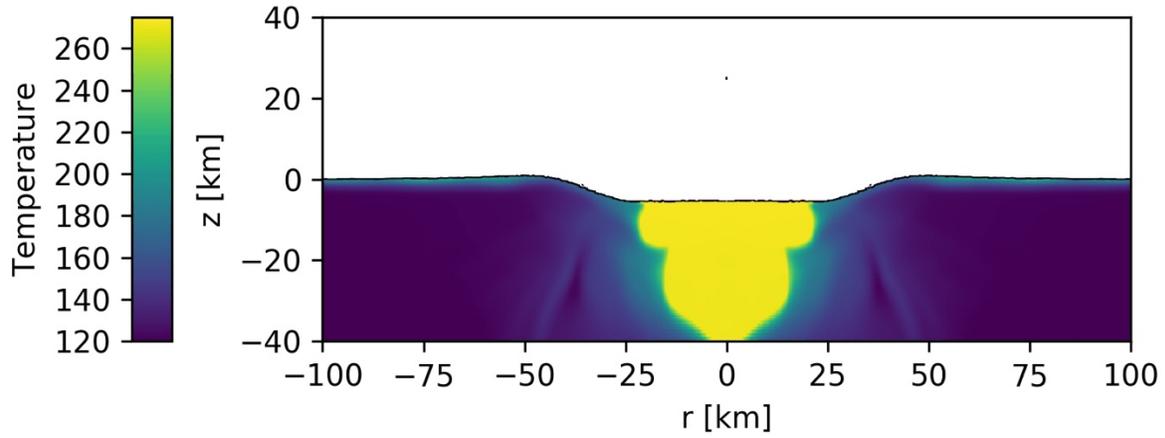

Figure E: Plot of temperature (left) and material b) for the default impact.

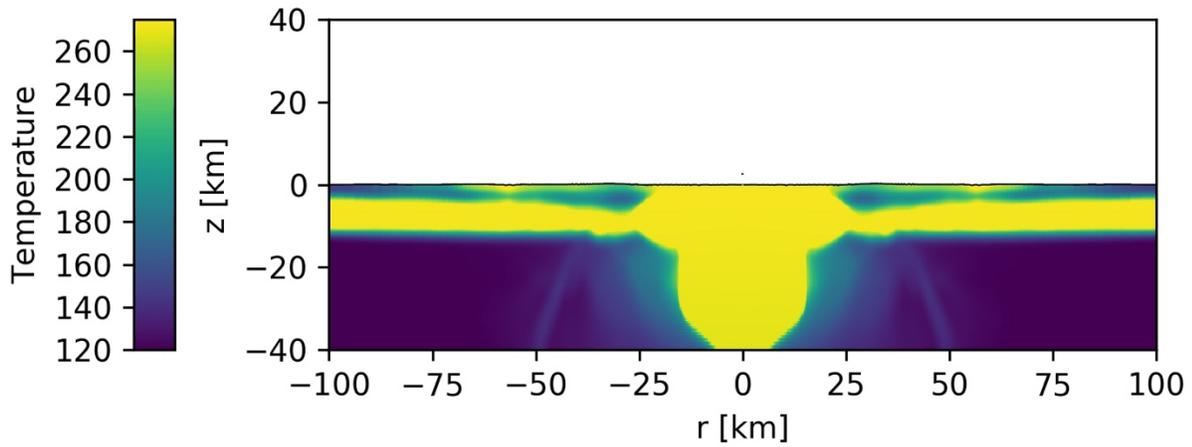

Figure F: Plot of temperature (left) and material b) for impact into target with pre-existing liquid layer −5 < $z$ < 10 km below the surface.

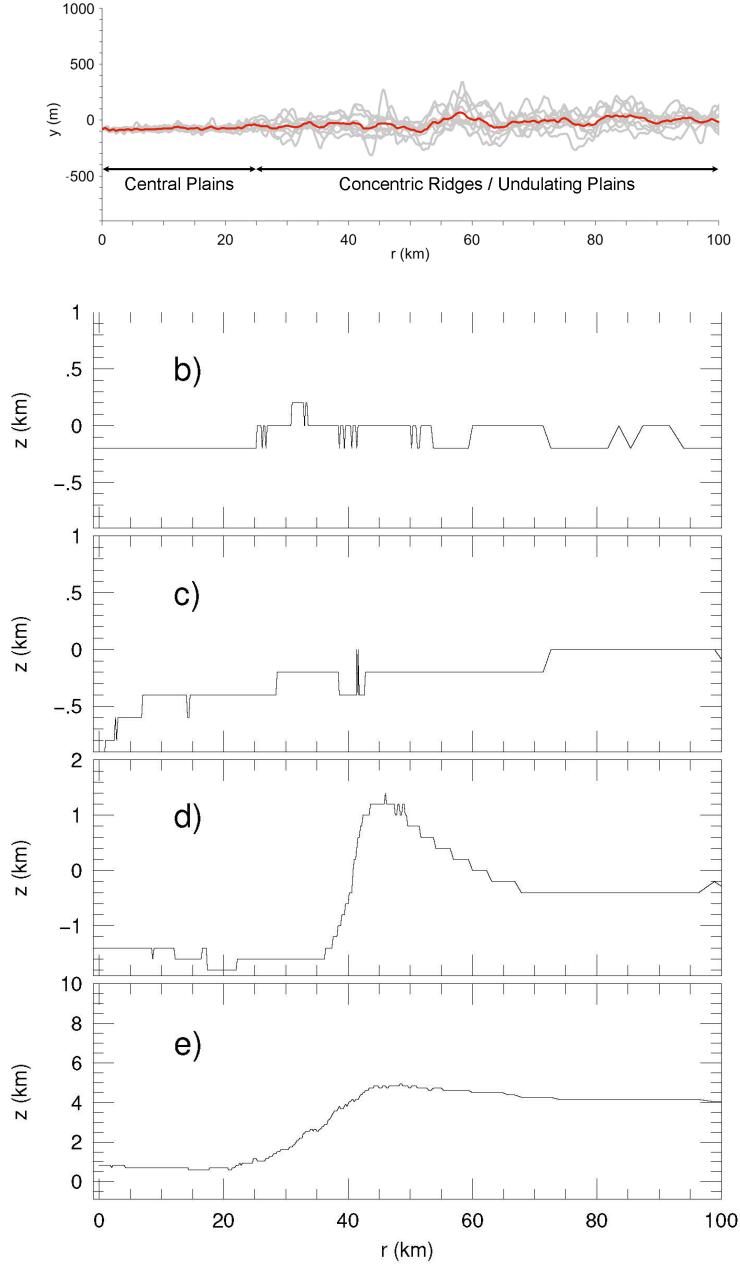

Figure G: (a) Ten averaged radial topographic profiles (shown in gray) measured across Buto crater out to 100 km radial distance from the center of the palimpsest. Averaged profile is shown in red and arrows indicate the radial expanses of the Central Plains and Concentric Ridges/Undulating Plains. (b-e) Surface profiles for simulated impacts ($d_i$ = 7.2 km, $U_i$ = 20 km s$^{-1}$) with pre-existing liquid layers (from top) $-10 < z < 5$ km (b), $-20 < z < 15$ km (c), $-40 < z < 35$ km (d), and $-60 < z < 55$ km (e) below the surface. Lateral scales of profiles in all figure elements are identical, while the vertical scales for profiles (a), (b), and (c) are identical.

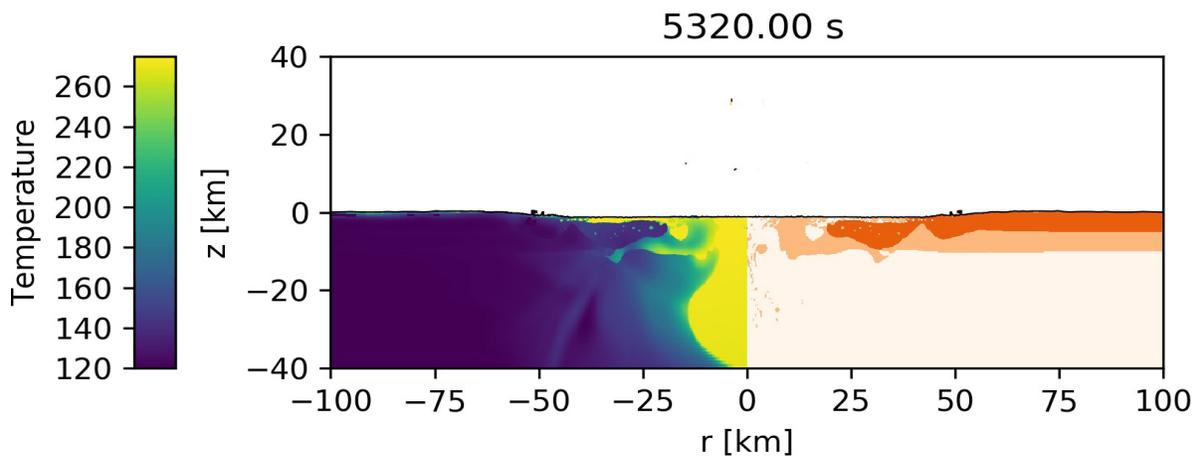

Figure H: Plot of temperature (left) and material b) for impact into target with weak layer $-5 < z < 10$ km below the surface

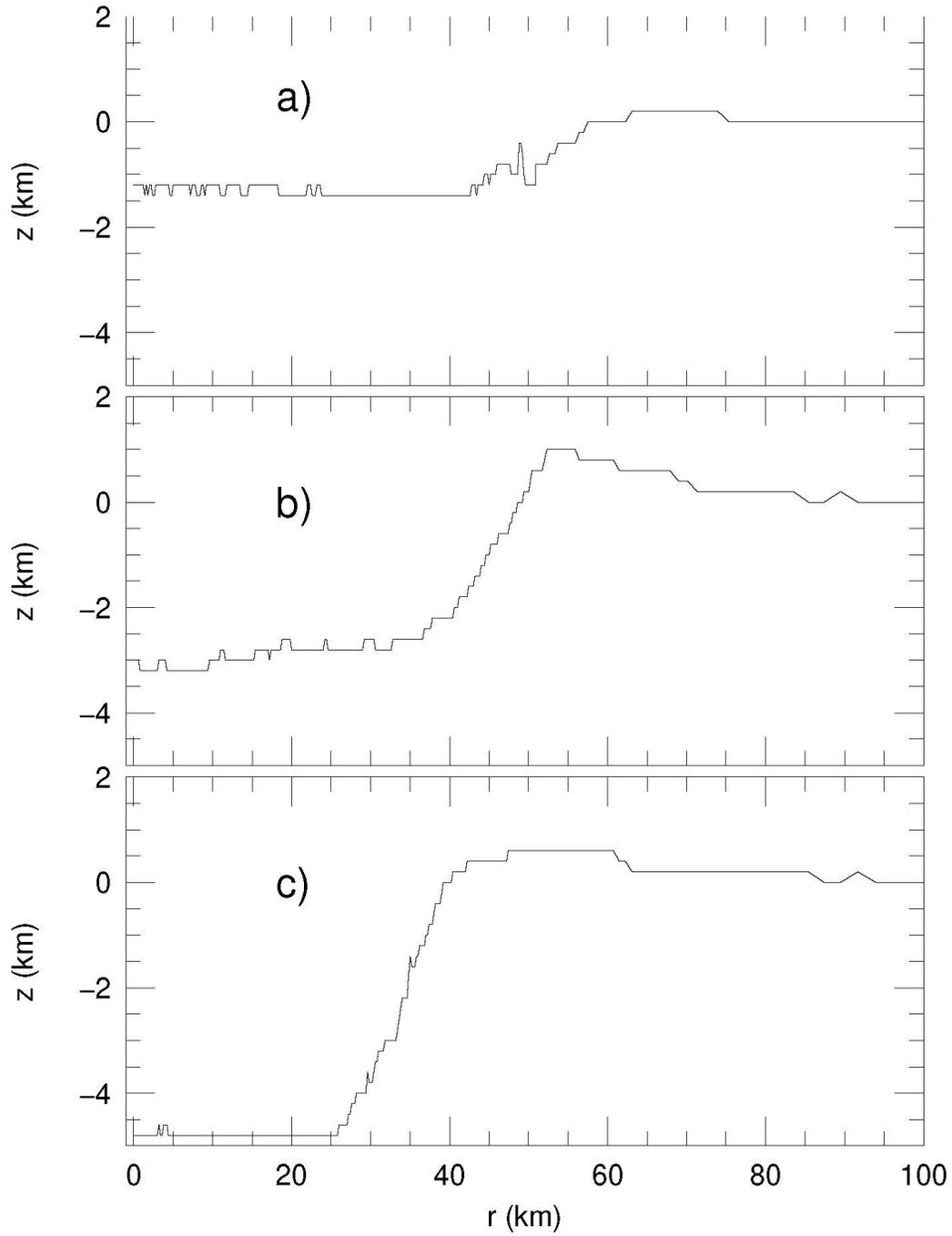

Figure I: Surface profiles for impacts ($d_i$ = 7.2 km, $U_i$ = 20 km s$^{-1}$) with weak layers (from top) $-10 < z < 5$ km (a), $-20 < z < 15$ km (b), and $-40 < z < 35$ km (c) below the surface.

# Table 1

| parameter | definition | value |
|---|---|---|
| **Intact ice:** | | |
| YINT0 | Cohesion | $10^7$ kg m$^2$ s$^{-2}$ |
| YLIMINT | Limiting strength | $10^8$ kg m$^2$ s$^{-2}$ |
| FRICINHT | Coefficient of internal friction | 0.6 |
| **Damaged ice:** | | |
| YDAM0 | Cohesion | $10^5$ kg m$^2$ s$^{-2}$ |
| YLIMDAM | Limiting strength | $10^8$ kg m$^2$ s$^{-2}$ |
| FRICDAM | Coefficient of internal friction | 0.6 |
| **Ivanov parameters:** | | |
| IVANOV_A | Minimum failure strain at low pressure | $10^{-4}$ |
| IVANOV_B | Failure coefficient | $10^{-11}$ |
| IVANOV_C | Compressional failure pressure | $10^8$ kg m$^2$ s$^{-1}$ |